\documentclass[aps,twocolumn,showpacs,amsmath,amssymb]{revtex4}
\usepackage{graphicx}
\usepackage{dcolumn}
\usepackage{bm}

\begin{document}

\title{Novel Extrapolation Method in the Monte Carlo Shell Model}
\author{Noritaka Shimizu$^1$}
\author{Yutaka Utsuno$^2$}
\author{Takahiro Mizusaki$^3$}
\author{Takaharu Otsuka$^{1,4,5}$}
\author{Takashi Abe$^1$}
\author{Michio Honma$^6$}
\affiliation{$^1$  Department of Physics, University of Tokyo, Hongo,
  Tokyo 113-0033, Japan}
\affiliation{$^2$ Advanced Science Research Center, Japan Atomic Energy Agency, Tokai,
  Ibaraki 319-1195, Japan }
\affiliation{$^3$ Institute of Natural Sciences, Senshu University, Tokyo, 
  101-8425, Japan}
\affiliation{$^4$ Center for Nuclear Study, University of Tokyo, 
  Hongo Tokyo 113-0033, Japan }
\affiliation{$^5$ National Superconducting Cyclotron Laboratory,
  Michigan State University, East Lansing, Michigan, USA}
\affiliation{$^6$ Center for Mathematical Sciences, Aizu University, Ikki-machi,
  Aizu-Wakamatsu, Fukushima 965-8580, Japan}

\date{\today}

\begin{abstract}
  We propose an extrapolation method utilizing energy variance 
  in the Monte Carlo shell model 
  in order to estimate the energy eigenvalue and observables accurately.
  We derive a formula for the energy variance with deformed Slater determinants,
  which enables us to calculate the energy variance efficiently.
  The feasibility of the method is demonstrated 
  for the full $pf$-shell calculation of $^{56}$Ni,  
  and the applicability of the method to a system beyond current 
  limit of exact diagonalization 
  is shown for the $pf$+$g_{9/2}$-shell calculation of $^{64}$Ge. 
\end{abstract}


\maketitle

The shell model (SM) calculation has been 
very successful in understanding the nuclear structure 
on the basis of nucleons interacting via the nuclear force.
The conventional, standard solver for SM calculations 
is the exact diagonalization of Hamiltonian matrix in a given model space.
Recently, the SM calculation plays an indispensable role especially in studying
neutron-rich exotic nuclei, including beta-decay properties on $r$-process nuclei
({\it e.g.} \cite{ppnp_brown, langanke_npa}). 
For such studies, the model space of the SM calculation 
should contain some intruder orbits in addition to one major shell.
In this case, the dimension of its Hilbert space is often 
explosively large and the practical calculation is infeasible.
Overcoming such a difficulty is a crucial challenge for modern SM calculations, 
where much effort has already been directed 
({\it e.g.} \cite{ppnp_mcsm, ppnp_schmid, hybrid_h2, horoi_pci, roth_importance}). 
The Monte Carlo shell model (MCSM) \cite{ppnp_mcsm} is one of the methods 
which aim at surpassing the limit of the conventional diagonalization \cite{rmp_caurier} 
and have succeeded in realistic applications.

The MCSM has been formulated by combining auxiliary-field quantum Monte Carlo 
and diagonalization methods \cite{mcsm_1995}. 
The MCSM yields the resulting  wave function 
as a linear combination of a relatively small number of 
deformed-basis wave functions.
While the convergence pattern of the energy eigenvalue 
as a function of the basis number 
suggests the validity of the approximation, 
the convergence is, in many cases, not fast enough to estimate the exact 
energies accurately.
This is a long-standing problem in the MCSM.
The same problem also occurs in the conventional SM calculations
when the model space is truncated.

In the case of the conventional SM calculations with truncation, 
the approximated eigenvalue
seems to decrease exponentially as a function of the basis number.
As an empirical trial, 
the exact energy can be guessed 
by an exponential extrapolation \cite{horoi_conv}, 
though this technique cannot be applied directly to the MCSM.  
In this paper, to estimate the exact energy eigenvalue, we consider another 
novel method free from such convergence patterns. 

Recently an extrapolation method utilizing energy variance to estimate exact
energy eigenvalue has been developed \cite{imada_pirg}.
Because this method is expected to be valid independently of the representation 
of the basis function, its application to the SM is of interest.
In spite of efforts for such applications \cite{extrap_2ndorder, extrap_slater}, 
its full-scale application has been 
infeasible due to the limitation of computer resources.
In the present work, by deriving a new formula for the expectation value of 
the Hamiltonian squared, such an extrapolation is made feasible.

First, we briefly review the framework of the MCSM.
We use a general two-body interaction as: 
\begin{equation}
  H = \sum_{ij} t_{ij} c^\dagger_i c_j 
  + \sum_{i<j, k<l} v_{ijkl} c^\dagger_i c^\dagger_j c_l c_k ,
\end{equation}
where $c^\dagger_i$ denotes a creation operator of 
single particle state $i$. 
In the present work, the MCSM wave function is given as a linear combination of 
angular-momentum-projected, parity-projected deformed Slater determinant wave functions, 
\begin{equation}
  |\Psi_N \rangle = \sum_{n=1}^N \sum_{K=-J}^J f^{(N)}_{n,K} P^{J\pi}_{MK}
  | \psi_n \rangle , 
\end{equation}
where $P^{J\pi}_{MK}$ is the angular-momentum 
and parity projector, 
and $N$ is called the MCSM dimension. 
Each $|\psi_n\rangle$ is a deformed Slater determinant, 
\begin{equation}
  |\psi_n \rangle = \prod_k \left( \sum_l D^{(n)}_{lk} c^\dagger_l \right) | - \rangle , 
\end{equation}
where 
$|-\rangle$ denotes an inert core.
The coefficient $D^{(n)}$ 
is selected from many (roughly one thousand) candidates 
generated stochastically utilizing the auxiliary field Monte Carlo technique.
The coefficient $f^{(N)}_{n,K}$ is determined 
by the diagonalization of the Hamiltonian matrix in the subspace spanned 
by projected Slater determinants, $P^{J\pi}_{MK} | \psi_n \rangle $.
This diagonalization also determines 
the energy, $E_N \equiv \langle \Psi_N | H |\Psi_N \rangle$, 
as a function of $N$.
In principle, we increase $N$ until $E_N$ becomes converged.

Next, we introduce the energy-variance extrapolation
into the MCSM (MCSM-extrapolation method) 
following the idea of Ref. \cite{extrap_2ndorder}.
The MCSM provides us with a successive sequence of the 
wave functions 
$|\Psi_1\rangle, |\Psi_2\rangle, \cdots , |\Psi_N\rangle, \cdots$. 
For each $N$, we evaluate energy variance as, 
\begin{equation}
  \langle \Delta H^2\rangle_N \equiv \langle \Psi_N| H^2|\Psi_N\rangle 
  - \langle \Psi_N | H |\Psi_N \rangle ^2 , 
  \label{eq:hsq}
\end{equation}
and plot the energy $E_N$ as a function of its variance.
As we increase $N$ and improve the approximation, 
the resulting energy approaches the exact energy, 
and the corresponding energy variance approaches zero.
These values are fitted by a second order polynomial, and 
the energy is extrapolated to the limit of 
$\langle \Delta H^2\rangle \rightarrow 0$
in the same manner as other applications of energy-variance extrapolation 
\cite{extrap_2ndorder}.

The obstacle in the implementation of the MCSM-extrapolation method 
was the large amount of computation to evaluate $\langle \phi |H^2| \psi\rangle$, 
where $|\phi\rangle$ and $|\psi\rangle$ are deformed Slater determinants.
If we regard $H^2$ as a general four-body operator, 
the evaluation of the matrix element consists of 
the eightfold-loop summation 
of the 24 terms of products of four generalized one-body density matrices, 
$\rho_{ij} = \langle \phi | c^\dagger_j c_i | \psi \rangle / \langle \phi | \psi \rangle$.
In the present work, 
thanks to the separability of $H^2$, the evaluation
of the matrix element is formulated as: 
\begin{widetext}
\begin{eqnarray}
  \frac{ \langle \phi | H^2 | \psi \rangle }
  { \langle \phi | \psi \rangle } 
  &=& 
  \sum_{i<j, \alpha < \beta}
  \left(
    \sum_{k<l} v_{ijkl}
    ( (1- \rho)_{k \alpha}(1-\rho)_{l \beta}  
    - (1- \rho)_{l \alpha}(1-\rho)_{k \beta} )
  \right)
  \left(\sum_{\gamma < \delta} 
    v_{\alpha \beta \gamma \delta} 
    ( \rho_{\gamma i} \rho_{\delta j} 
    - \rho_{\delta i} \rho_{\gamma j} )
  \right) 
  \nonumber \\
  && + {\rm Tr}((t+\Gamma) (1-\rho)(t+\Gamma) \rho)
  + \left( {\rm Tr}(\rho(t+\frac12 \Gamma))\right)^2 
  \label{eq:two-two}
\end{eqnarray}
\end{widetext}
with $\Gamma_{ik} = \sum_{jl} v_{ijkl} \rho_{lj}$.
The trivial summations and their indices 
for the matrix products are omitted.
The first term in Eq.(\ref{eq:two-two}) is written as a product of two matrices 
as the first term on the right-hand side. 
This factorization reduces the eightfold loop 
into a sixfold loop and decreases the computation time drastically.

Now, we apply the MCSM-extrapolation method 
to $^{56}$Ni with the $pf$-shell and the FPD6 interaction \cite{fpd6}. 
The $m$-scheme dimension of $^{56}$Ni reaches $1.0\times 10^9$.
The present work was performed 
using the newly developed MCSM code \cite{riken_rmcsm}, 
which enables us to run it on latest supercomputers.

\begin{figure}[ctb!]
  \begin{center}
    \includegraphics[width=7.0cm]{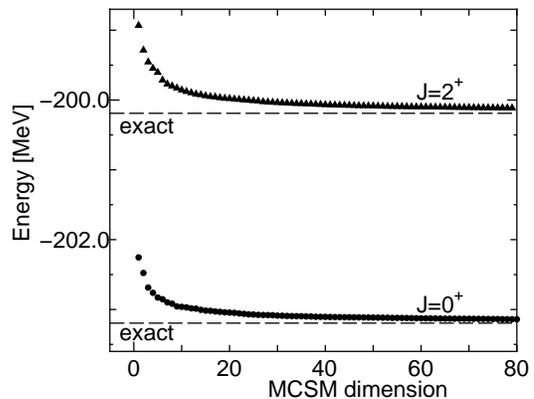}
    \caption{ Convergence patterns of the ground  
      and first excited states of $^{56}$Ni in the $pf$-shell.
      The solid circle and the triangular symbols denote 
      the MCSM results of $J=0^+$ and  $J=2^+$, respectively.
      The dashed lines show the exact values by the diagonalization method.
    }
    \label{fig:ni56_conv}
  \end{center}
\end{figure}

Figure \ref{fig:ni56_conv} shows the MCSM results of 
the ground-state ($J^\pi=0^+$) 
and the first-excited-state ($J^\pi=2^+$) energies 
as functions of the MCSM dimension.
These energies show good convergences,
but slight differences from the exact values remain.
We will show how these gaps are filled by the extrapolation method later.
The energy by the current MCSM calculation is $-203.161$ MeV with $N=150$, 
while the past results of the MCSM were $-203.100$ MeV in 1998 \cite{mcsm1998}, 
and $-203.152$ MeV in 2001 \cite{ppnp_mcsm}.
Over a decade, progress in the method and 
in computational power 
has gradually improved the precision of the MCSM.
Nevertheless, we still find $37$ keV error from the exact energy, 
$-203.198$ MeV. 
Note that the MCSM error of the $2^+_1$ state is the same order of magnitude.

\begin{figure}[ctb!]
  \begin{center}
    \includegraphics[width=8.0cm]{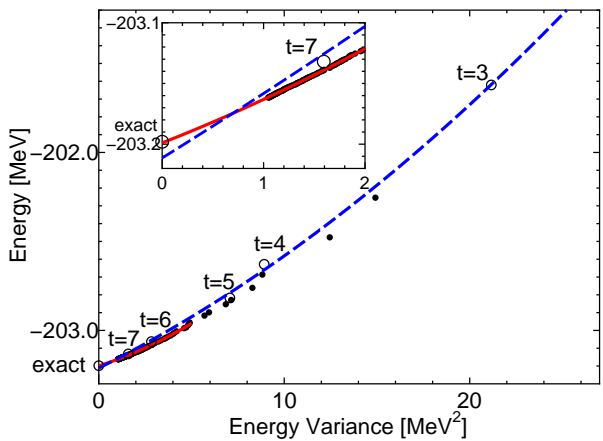}
    \caption{ (Color online) Second-order extrapolations of the ground-state 
      energy  into a zero energy variance of the 
      $J^\pi=0^+$ ground state of  $^{56}$Ni in the $pf$-shell.
      The filled symbols, open symbols, solid red line, and dotted blue line 
      denote the $E_N$ of MCSM, 
      the results of the diagonalization method with PHT,
      and their second-order fits, respectively. 
      The exact energy is also shown by open symbols 
      on the $y$-axis.
      The inset shows  magnified view around $\langle \Delta H^2\rangle \simeq 0$. 
    } 
    \label{fig:ni56_evariance}
  \end{center}
\end{figure}

Figure \ref{fig:ni56_evariance} shows the $E_N$ as a function of 
$\langle \Delta H^2\rangle_N $ 
provided by the MCSM wave function.
We fit the MCSM points of $E_N$ against $\langle \Delta H^2\rangle_N $ 
with $10\leq N\leq 150$ by quadratic curve, 
and extrapolate the MCSM results to 
$\langle \Delta H^2\rangle \rightarrow 0$.
The extrapolated energy is $-203.198$ MeV,
which agrees with the exact one within $1$ keV. 
Here, we excluded the first nine points of $E_N$ 
for the quadratic fit because the extrapolation method assumes that 
approximated wave functions are sufficiently close to 
the true eigenstate.
Moreover, the MCSM points of $N<10$ show
comparably large fluctuation due to stochastic procedure
and should have strong dependence on the initial states of 
stochastic sampling. 

For comparison, we also show another extrapolation result 
for the conventional SM calculation 
with the particle-hole truncation (PHT) in Fig.\ \ref{fig:ni56_evariance}.
The configuration of the PHT is 
$(0f_{7/2})^{16-t} (0f_{5/2},1p_{3/2},1p_{1/2})^{t}$
with $t=3,4,5,6,7$, and the practical calculation 
was performed by the MSHELL code \cite{mshell}.
These energies and their variances are also fitted 
by a quadratic curve in the same manner as Ref. \cite{extrap_2ndorder}.
While both the MCSM and the PHT calculation 
succeed in reproducing the exact energy well, 
minor deviation can be seen in the inset of Fig. \ref{fig:ni56_evariance}.
The extrapolated energy with PHT
is $-203.217$ MeV, and its discrepancy with the MCSM and the exact energy 
is $19$ keV.
Note that we discuss precision in the unit of a few keVs, 
while previous works using the energy-variance extrapolation 
provided the precision of a few tens or a hundred keVs 
\cite{extrap_2ndorder, extrap_slater}.

An advantage of the MCSM for the extrapolation method 
is that the MCSM provides us with the sequence of 
many (more than 50) successive approximate wave functions simultaneously.
It provides us with good statistics for the extrapolation.
On the other hand, the conventional PHT scheme 
yields only 6 points in the case of $^{56}$Ni, for example.

In order to test the applicability to larger systems,
we assume that the MCSM result with $N\leq 50$ is available 
in the ground state of the $^{56}$Ni case.
In practical calculations, the $N$ is often limited so small 
that the $E_N$ cannot reach good convergence.
The MCSM result with $N=50$ is $E_{N=50} = -203.115$ MeV, 
which is worse than $t=7$ energy, $-203.132$ MeV.
Nevertheless, the extrapolated energy of the MCSM is $-203.202$ MeV, which 
is still much closer to the exact result
than that of PHT.
This good agreement 
provides us with a promising perspective for its application 
to larger systems.

\begin{figure}[ctb!]
  \begin{center}
    \includegraphics[width=7.0cm]{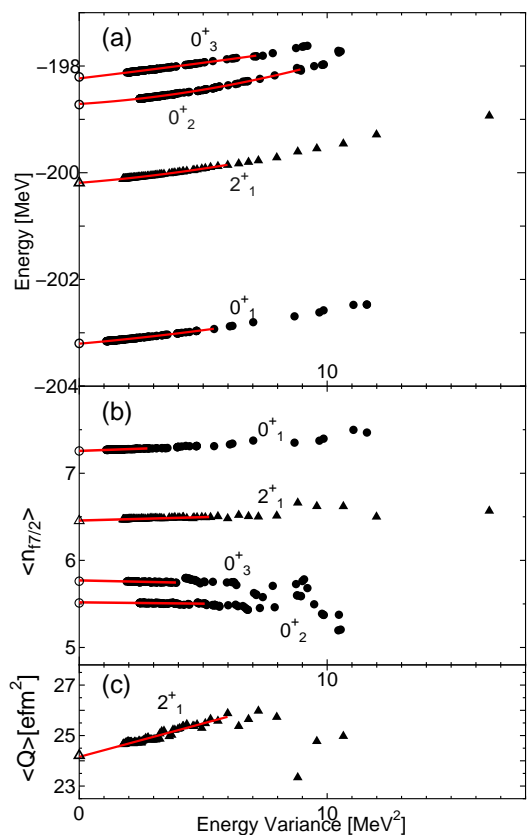}
    \caption{ (Color online) (a) Second-order extrapolations of the energies
      of $J^\pi=0^+_1, 2^+_1, 0^+_2,$ and  $0^+_3$ states of  
      $^{56}$Ni in the $pf$-shell.
      The notation is the same as Fig. \ref{fig:ni56_evariance}.
      (b)First-order extrapolation of the occupation number of 
      the $0f_{7/2}$ orbit by the MCSM. 
      (c) First-order extrapolation of the quadrupole moment of 
      the $2^+_1$ state.
      The results obtained by exact diagonalization are also shown by 
      the corresponding open symbols. } 
    \label{fig:ni56_excited}
  \end{center}
\end{figure}

Figure \ref{fig:ni56_excited} shows 
the results of $0_1^+, 2_1^+, 0_2^+,$ and  $0_3^+$ states in order to discuss 
the behavior of the MCSM extrapolation concerning excited states 
and some observables.
In Fig.\ \ref{fig:ni56_excited}(a), all of the MCSM-extrapolation results of 
these energies agree excellently with the exact ones in a unit of keV, too.
Figures \ref{fig:ni56_excited}(b) and (c) show
the occupation numbers of  the $0f_{7/2}$ orbit 
and the quadrupole moment of the $2^+_1$ state by the MCSM 
and their first-order extrapolations.
In the case of these observables, 
a first order polynomial is appropriate for the extrapolation
because the positive and negative 
contributions of the contamination of excited states cancel each other.
Obviously, such cancellation does not occur in the case of energy eigenvalue.
The first-order extrapolation for these observables provides us with 
excellent improvement of the agreement with the exact value, 
while some other extrapolation methods do not 
\cite{extrap_slater}.

\begin{figure}[ctb!]
\begin{center}
  \includegraphics[width=7cm]{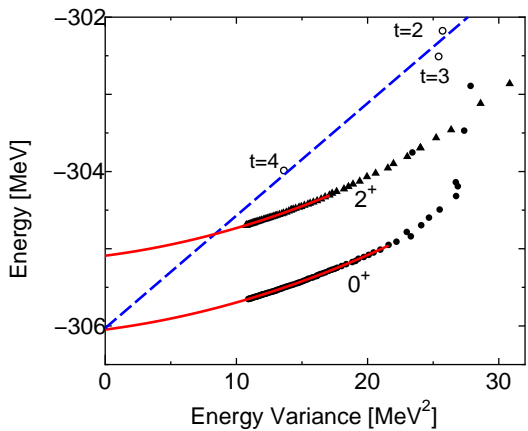}
\caption{ (Color online) Second-order extrapolations of the ground-state 
  and $2^+$ energies of $^{64}$Ge in the $pf$+$g_{9/2}$-shell.
  The blue dashed line shows the first-order extrapolation of ground-state energy 
  of PHT calculation.
  The notation is the same as that of Fig. \ref{fig:ni56_evariance}.
}
\label{fig:ge64_h2}
\end{center}
\end{figure}

Finally, we discuss the case of $^{64}$Ge with $pf+g_{9/2}$ model space
in order to demonstrate the applicability of the present method 
to large-scale SM calculations. Its $m$-scheme dimension is 
$1.7\times10^{14}$, which is roughly $10^3$ times larger than the current limitation of 
the conventional diagonalization method, $\sim 10^{11}$.
We adopt the PFG9B3 effective interaction \cite{pfg9b3}, 
which was used also in Ref.\cite{horoi_leveldens}.
In Fig.\ \ref{fig:ge64_h2}, the result of the MCSM-extrapolation method 
shows stable behavior while the exact value is not available.
The 82 points for the ground state are obtained by the MCSM 
and fitted by a quadratic curve.
The excitation energy of $2^+$ state is $0.95$ MeV, 
which is close to the experimental value, $0.90$ MeV \cite{nudat}.
We also see the reasonable agreement between the ground-state energy of MCSM-extrapolation 
and that of the first-order extrapolation with PHT calculation.
We point out that the PHT extrapolation is based on the four points ($1\leq t \leq 4$), 
and the fitted line shows certain deviations from these points already, 
suggesting possible ambiguities.
Note that the guess by the statistics of the nuclear level density 
is rather low, $-306.7$ MeV   \cite{horoi_leveldens}.

In summary, we have proposed the MCSM-extrapolation method 
which provides us with accurate correction to the MCSM.
Eq. (\ref{eq:two-two}) considerably reduces
the computation time by orders of magnitudes 
to calculate the energy variance with deformed Slater determinants.
The energy as a function of its variance 
is well fitted by a quadratic curve, and 
the result of the MCSM is improved down to a unit of keV especially in $^{56}$Ni case.
We demonstrate that this method works quite well 
not only for energy eigenvalues, 
but also for other physical quantities of some low-lying states.
By adopting the extrapolation method with the energy variance, 
we obtain a self-contained framework which removes the ambiguity 
of the energy convergence in the MCSM.
We applied this framework also to 
large-scale shell-model problems, like the case of $^{64}$Ge, 
which cannot be solved by existing conventional solvers.
These results look quite promissing and encourages us 
to apply the present method to larger-scale problems.
In such cases, the error estimation of the extrapolation method itself 
becomes important, and will be discussed in future publication.

We thank Prof. M. Horoi for valuable discussions.
This work has been supported by Grants-in-Aid
for Young Scientists (20740127), (21740204), 
for Scientific Research (20244022) 
and for Scientific Research on Innovative Areas (20105003) 
from JSPS. 
It has also been supported by 
the CNS-RIKEN joint project for large-scale nuclear structure calculations. 
A part of the numerical calculation was performed 
on the T2K Open Supercomputer at the University of Tokyo and the BX900 Supercomputer
at Japan Atomic Energy Agency.


\end{document}